\documentstyle{article}[12pt]
\begin{document}
\def\et{\epsilon _3}
\def\es{\epsilon _4}
\def\w{\wedge}
\def\be{\begin{equation}}
\def\ee{\end{equation}}
\def\pv{{\bf P}(V)}
\def\a{\alpha}

\vskip 1cm
\begin{flushright}
 solv-int/9608006
\end{flushright}
\vskip 1cm
\begin{center}

 {\large On Simplest Hamiltonian Systems}
\vskip 0.5cm
Pavol \v Severa
\vskip 0.25cm
{\small \it Dept. of Theoretical Physics, Charles University\\
V Hole\v sovi\v ck\' ach 2 182 00 Prague, Czech Republic}
\vskip 1cm

\begin{abstract}
 Simple Hamiltonian systems, such as mathematical pendulum or Euler
equations for rigid body, are solved without computation. It is nothing
but a joke but maybe you will find it nice.
\end{abstract}
\end{center}
\vskip 0.75cm
There are some simple mechanical systems, such as mathematical pendulum
or Euler equations for rigid body, admitting solution in elliptic functions.
These functions are one of the most beautiful things in elemenary complex
analysis, especially when they are treated in Riemannian way, using their
global properties and singularities and avoiding computation. As a simple,
but maybe charming example, we present a geometrical solution of the
afore-mentioned systems.

Let our phase space be an irreducible quadratic surface $F=0$ in a
3-dimensional affine space $A_3$. The symplectic form is given as follows:
choose
a trivector $\et$; then $\sigma =<\et ,dF>$ (when restricted to $F=0$)
is the bivector field on the phase space giving the Poisson structure
(i.e. the inverse of the symplectic form). It is invariant with respect to
unimodular affine transformations preserving $F$ and, if all singularities of 
$F=0$ are hidden at infinity, this property fixes $\sigma$ up to a constant 
multiple. If $H$ is the Hamiltonian, then
\be v=<\et ,dF \w dH> \ee
is the evolution field.

Suppose that $H$ is a time-independent quadratic function on $A_3$. The best
known examples are Euler equations (the phase space a sphere) and
(possibly rotating) mathematical pendulum (the phase space a cylinder).
Any energy level, say $H=0$, is a spatial quartic curve. As soon as it is
regular, it is an elliptic curve (if we project it from its arbitrary point,
we obtain a regular planar cubic). We will see that (1) is regular on the
curve, including the complex points and the poins at infinity. Thus the 
evolution is nothing but addition on the elliptic curve and the system is
solved in elliptic functions.

We will restate (1) using projective geometry. Let $V$ be a vector space,
$C$ a 2-dimensional cone in $V$ and $c$ the corresponding curve in the 
projective space $\pv$. A vector field $v$ on $c$ can be realized in $V$ as
a bivector field $b$ on $C$, homogenous of degree 2. The correspondence is
as follows:

Let $P\in C$ and let $p\in c$ be the corresponding point in $\pv$ and let
$v'$ be a vector at $P$ projected to $v$ at $p$ ($v'$ is unique up to
addition of a constant multiple of $P$). Then $b=P\w v'$. Also, if
$\a\in V^*$ is such that $\a (P)=1$ then $<b,\a >$ is a possible choice
of $v'$.

Now let $A_3$ be realized in a 4-dimensional $V$ by an equation $\a=1$.
Also let $f,h$ be the homogenizations of $F,H$. Let $\es$ be the unique 
tetravector on $V$ such that $\et=<\es,\a>$. Then, according to (1),
on the cone $f=h=0$,
\be b=<\es,df\w dh>. \ee
Equation (2) is a priori valid on $\a=1$, but in fact it is valid everywhere
on $C$, as $df\w dh$ has the right homogenity.

We see that $b$, and thus $v$, is everywhere finite.

If $df\w dh=0$ anywhere on the curve, the situation is even more favourable.
The curve is necessarily singular (on an elliptic curve, $v$ is nowhere zero)
and thus it is a rational curve or a reducible curve composed of rational ones.
All the following assertions are obtained by considering the curve as a 
limiting case of a regular one.

If the curve is irreducible with a cusp, equations of motion are solved in
rational functions: if the cusp is identified with $\infty$, the evolution
is simply addition.

If the curve has a simple double point, solution can be given in exponentials:
if the coalescing points are identified with $0$ and $\infty$, the evolution
is multiplication. The exponential is real or imaginary according to the
location of the double point with respect to the other real points.

If the curve decomposes (to a line and a twisted cubic, or to two
(possibly reducible) conics), the evolution is rational or exponential
according to the location of points of intersection. If a component is double,
$df\w dh=0$ identically there and the evolution is trivial.

\end{document}